\documentclass[a4paper,11pt]{article}

\usepackage{jcappub} 

\usepackage[T1]{fontenc} 
\usepackage{graphicx}
\usepackage[normalem]{ulem}

\title{\boldmath Evolution of Vacuum Bubbles Embedded in Inhomogeneous Spacetimes}

\author[a,1,2]{Florencia Anabella Teppa Pannia,\note{Fellow of CONICET.}\note{Corresponding author.}}
\author[b]{Santiago Esteban Perez Bergliaffa,}

\affiliation[a]{Grupo de Astrof\'{\i}sica, Relatividad y Cosmolog\'{\i}a, Facultad de Ciencias Astron\'omicas y Geof\'{\i}sicas, Universidad Nacional de La Plata, Paseo del Bosque s/n B1900FWA, La Plata, Argentina.}
\affiliation[b]{Departamento de F\'{\i}sica Te\'{o}rica,
Instituto de F\'{\i}sica, Universidade do Estado de Rio de Janeiro, CEP 20550-013, Rio de Janeiro, Brazil.}

\emailAdd{fteppa@fcaglp.unlp.edu.ar}
\emailAdd{sepbergliaffa@gmail.com}

\abstract{
  We study the propagation of bubbles of new vacuum in a radially inhomogeneous background filled with dust or radiation, and including a cosmological constant, as a first step in the analysis of the influence of inhomogeneities in the evolution of an inflating region. We also compare the cases with dust and radiation backgrounds and show that the evolution of the bubble in radiation environments is notably different from that in the corresponding dust cases, both for homogeneous and inhomogeneous ambients, leading to appreciable differences in the evolution of the proper radius of the bubble.
}  

\begin{document} 
\maketitle
\flushbottom

\section{Introduction}
\label{sec:intro}
An inflationary phase can solve some of the issues present in the standard cosmological model (such as the horizon and flatness problems).  However it is not clear to what extent this assertion depends on assuming that the pre-inflationary universe is described by a homogeneous and isotropic spacetime. 
 In particular, the onset of inflation in the presence of inhomogeneities produced by a pre-inflationary era, possibly driven by quantum gravity, has been discussed by several authors.  A pioneer work on the analysis of de Sitter space-time as a late-time attractor for universes with generic initial conditions was presented in \cite{Starobinski1983}, where it was shown that a non-zero effective cosmological constant can smooth-down all types of inhomogeneities, even in anisotropic geometries. The attractor property of power-law inflation for inhomogeneous cosmological models was also discussed in \cite{Muller1990}.  Numerical studies were presented in \citep{Piran1988,Goldwirth1989b,Goldwirth1989,Goldwirth1990,Goldwirth1992,Deruelle1995,Perez2011}, among others, where it was claimed that highly homogeneous and isotropic initial conditions on a patch several times larger than the horizon are necessary for inflation to start. However other authors \citep{Kurki1993,Berera2001,Clough2016,Brandenberger2016} asseverate that inflation is also viable with inhomogeneous initial conditions. For recent studies on this subject see \cite{East2016,Kleban2016}, where numerical evolution of the full non-linear Einstein's equations for an inflaton field coupled to gravity was computed for different types of inhomogeneities. 

Once inflation has started in a certain region, it remains to see whether the ambient inhomogeneities affect its development  and, in particular, if inflation will effectively smooth out large initial inhomogeneities. These inquiries are relevant not only in the context of traditional early-universe models of inflation, but also in the more speculative scenario of eternal inflation and the string landscape, in which regions filled with new vacuum nucleate into an ambient region, leading to inflationary patches in different environments.\footnote{Initial conditions must be handled with care in this case, see discussion in Sect.~\ref{sec:discussion}.}  Since the nucleation does not necessarily occur in vacuum-dominated regions, it is interesting to study the time development of such patches in less symmetric ambients with different matter contents.  For instance, dust inhomogeneous environments should be considered to study cases of eternal inflation, where the bubble would eventually meet an inhomogeneity in a
matter-dominated universe. Radiation inhomogeneous environments, however, are more appropriate to describe the ultra-relativistic matter present in the early universe.

One possibility to address this issue is to consider the so-called {\it thin-shell approximation}, which provides a simple treatment to described the evolution of two spacetime regions separated by a spherically-symmetric shell. This formalism, firstly developed by Israel \citep{Israel1966}, is based on the assumption of the continuity of the metric functions across the whole spacetime by taking into account appropriated junction conditions \cite{Berezin1987}, and was widely used to describe the evolution of nucleated bubbles of new vacuum patches in different scenarios regarding to vacuum-energy or dust bubbles immersed in de Sitter, Schwarzschild and FLRW  \citep{Maeda1983b,Maeda1983c,Laguna-Castillo1986,Sato1986,Lake1987,Ng2011}. 
 Concerning the evolution of shells in less symmetric scenarios, the growth of true vacuum bubbles embedded in an inhomogeneous spherically-symmetric background described by the Lema\^{\i}tre-Tolman-Bondi (LTB) solution of the Einstein's equations (see for instance \citep{Bolejko2010})
has been analysed in \cite{Fischler2008,Simon2009,Rakic2010}.\footnote{The problem regarding to aspherical perturbations and their consequences on the stability of vacuum bubbles was studied in \cite{Aguirre2005}. More recently, the study of spherically-symmetric embeddings of FLRW cosmological bubbles into various background spacetimes using the Raychaudhuri equation for null rays was presented in \cite{Shajidul2017}.} 
In this case the geometry is sourced by a pressureless fluid, which models inhomogeneities generated by a previous inflationary era. 

In this work we generalise the analysis in \cite{Fischler2008}, by replacing the  spherically-symmetric dust distribution in the ambient spacetime
by radiation with the same symmetries, the geometry being  characterised  by the FLRW and Lema\^{\i}tre's geometries  \cite{Lemaitre1933}, for homogeneous and inhomogeneous cases, respectively. This setting is appropriate to describe inhomogeneities in the early universe which
formed during a pre-inflationary era, as in  \cite{Goldwirth1989b,Goldwirth1990,East2016}, and are described by ultra-relativistic matter. 
 The evolution of the bubble is computed numerically, together with the evolution of the external metric, using the thin-shell formalism. 
 Our results show that the evolution in a radiation background is markedly different from that of a dust background.
  The differences are appreciable in the homogeneous and isotropic case, as well as in the inhomogeneous case, and lead to noticeable differences in the evolution of the proper radius of the bubble.
  
 The paper is organised as follows. In Section~\ref{sec:thin_shell} the thin-shell formalism is described, including a detailed characterisation of Lema\^{\i}tre's solution. In order to analyse the different effects of the matter content on the growth of the bubble, we start by comparing in Section~\ref{sec:PvoRad} the evolution of vacuum regions in dust and radiation homogeneous backgrounds. The inhomogeneous cases are studied in Section~\ref{sec:inhom_back}. We close with some remarks about the results in Section~\ref{sec:discussion}. 
 %
 %
 \section{The thin-shell formalism}
 \label{sec:thin_shell}
  We shall assume that an inflating vacuum patch is embedded in a generic environment.  The two spacetime regions, denoted here by ${\cal M}^-$ and ${\cal M}^+$, are separated by a time-like hypersurface $\Sigma$, which  has negligible thickness and its matter content is described by a given equation of state. The metric tensor is required to be continuous across the whole spacetime, and the total energy-momentum tensor is allowed to be discontinuous at the thin-shell, but it is continuous elsewhere. The junction conditions of the thin-shell formalism are devised to glue the two spacetimes in such a way that both geometries, as well as the shell that separates them, are a solution of Einstein's equations.
 Although it is assumed that the metric is continuous across the whole spacetime, jump discontinuities in the extrinsic curvature $K_{a b}$ are possible. This implies that the Einstein tensor (which involves second derivatives of the metric), and so the energy-momentum tensor $T_{a b}$, can have a jump discontinuity and/or a $\delta$-function singularity. We consider the field equations with non-vanishing cosmological constant, written as 
\begin{equation}
\label{EinsteinEqns}
G_{\mu\nu}= T_{\mu\nu} + \Lambda g_{\mu\nu}\,, 
\end{equation} 
where the units are chosen as $c=8\pi G=1$.
 In order to study the evolution of the shell, we need to solve these equations requiring the continuity condition for the metric through the surface layer.  The matching is such that both geometries evolve independently and the properties of the shell vary with time to adjust for local changes of the spacetime.
 We shall restrict our analysis to spherically-symmetric cases in which the inner vacuum region is described by the de Sitter metric, while the matter content of the outer region (composed by dust or radiation) may have an homogeneous or inhomogeneous distribution. 
  The set of general equations that solve Einstein's equations for an arbitrary hypersurface layer is carefully reviewed in \cite{Laguna-Castillo1986}.

The most general inhomogeneous spherically-symmetric solution of Einstein's equations with non-zero matter content is described by Lema\^{\i}tre's geometry, along with an equation of state of the form $p(t,r)=\lambda\epsilon(t,r)$.  The homogeneous FLRW solution and the inhomogeneous dust solution can be recovered, respectively, in the limits $\epsilon(t,r)=\epsilon(t)$ and $\lambda=0$. 
  Following the notation used  in \cite{Fischler2008}, the problem is then characterised as follows:\footnote{Hereafter, the subscripts ``-'' and ``+'' indicate, respectively, inner and outer quantities.}
\begin{itemize}
\item The inner vacuum region, ${\cal M}^-$, with non-zero cosmological constant $\Lambda^-$, is described by an isotropic and homogeneous metric. Using the coordinates ($T,z,\theta,\phi$), the line element is 
  \begin{equation}
     {\rm d}s^2|_{{\cal M}^-} ={\rm d}T^2-b^2(T)\left(\frac{{\rm d}z^2}{1+z^2}+z^2{\rm d}\Omega^2\right)\,, \\
  \end{equation}
where the evolution of the scale factor $b(T)$ is given by  
  \begin{equation}
  \left(\frac{{\rm d}b}{{\rm d}T}\right)^2=\left(\frac{\Lambda^{-}}{3}\right)b^2(T)+1\,.
  \end{equation}
\item  The spherically-symmetric time-like shell is characterised by the hypersurface $\Sigma$ with normal unit vector $n_\mu$ directed from ${\cal M}^-$ to ${\cal M}^+$ ($n^\mu n_\mu=-1$) and metric $h_{\mu\nu} =g_{\mu\nu}-n_{\mu}n_{\nu}$. The line element expressed in coordinates ($\tau,\theta,\phi$) on the bubble is
 \begin{equation}
     {\rm d}s^2|_{\Sigma} = {\rm d}\tau^2-\rho(\tau)^2{\rm d}\Omega^2\,.
  \end{equation}
 We assume that the matter content on the shell can be described by a perfect fluid with energy-momentum tensor given by 
  \begin{equation}
  S_{\mu\nu}=[\sigma(\tau) + \Pi(\tau)]v_{\mu}v_{\nu}-\Pi(\tau) h_{\mu\nu}\,,
  \end{equation}
 where $\sigma$ and $\Pi$ denote, respectively, the energy density and the pressure of the fluid on the bubble, and $v_\mu$ is the four-velocity of an observer on $\Sigma$.  Strictly, this tensor is defined in the thin-shell approximation as the integral of the effective energy-momentum tensor over the thickness of the hypersurface $\Sigma$ as the thickness goes to zero, that is,  
\begin{equation}
\label{Sintegral}
S_{\mu\nu} \equiv \lim_{y\rightarrow 0} \int^y_{-y} (T_{\mu\nu}+\Lambda g_{\mu\nu}) {\rm d}n = \lim_{y\rightarrow 0} \int^y_{-y} G_{\mu\nu} {\rm d}n\,,
\end{equation}
where $n$ is the proper distance through $\Sigma$ in the normal direction given by the orthogonal coordinate $y$, so that $y=0$ at $\Sigma$ \cite{Gron2007}.

\item The outer region, ${\cal M}^+$, is modelled by the Lema\^{\i}tre's solution \cite{Lemaitre1933}, with line element in external coordinates ($t,r,\theta,\phi$) given by 
  \begin{equation}
  \label{ds_Lem}
        {\rm d}s^2|_{{\cal M}^+} = {\rm e}^{A(t,r)}{\rm d}t^2-{\rm e}^{B(t,r)}{\rm d}r^2-R^2(t,r){\rm d}\Omega^2\,,
  \end{equation}
 and  energy-momentum tensor 
   \begin{equation}
  T_{\mu\nu}=[\epsilon(t,r) + p(t,r)]u_{\mu}^+u_{\nu}^+ -p(t,r) g_{\mu\nu}^+ \,. 
  \end{equation}
The Einstein's field equations (\ref{EinsteinEqns}) for the outer metric are
\begin{eqnarray}
  \label{EE1}
  R^2(t,r)R'(t,r)\epsilon(t,r)&=&2M'(t,r)\,,  \\
  \label{EE2}
 R^2(t,r)\dot{R}(t,r)p(t,r)&=&-2\dot{M}(t,r)\,,
\end{eqnarray}
 where the symbols $\ \dot \ $ and $\ '\ $ indicate, respectively, derivatives with respect to $t$ and $r$. The function  $M(t,r)$ satisfies the equation
\begin{equation}
\label{dRdt}
2M(t,r)=R(t,r)+{\rm e}^{-A(t,r)}\dot{R}^2(t,r)R(t,r) -{\rm e}^{-B(t,r)}R'^2(t,r)R(t,r)-\frac{\Lambda^+ R^3(t,r)}{3}\,,
\end{equation}
and the conservation of $T_{\mu\nu}$ yields the following relations:
\begin{eqnarray}
  \label{eq_A}
  A'(t,r)&=&-\frac{2p'(t,r)}{\epsilon(t,r)+p(t,r)}\,, \\
  \label{eq_B}
   {\rm e}^{B(t,r)}&=&\frac{R'^2(t,r)}{1+2E(r)}
      {\rm exp}\left(\int^t_{t_0}\frac{2\dot{R}p'}{[\epsilon+p]R'}{\rm d}\tilde t\right), 
\end{eqnarray}
 where $E(r)$ is an arbitrary function related to the local curvature and $t_0$ is the time corresponding to the initial conditions \citep{Bolejko2006}.  
\end{itemize}
 It is important to note that although the metrics are expressed in different coordinate systems, their angular coordinates coincide due to the spherical symmetry of the problem.

 Adopting the usual notation $[A]\equiv A^+-A^-$, the continuity condition for the metric through the hypersurface layer can be stated as $[g_{a b}]=0$. This condition imposes the following restrictions for the metric functions: 
\begin{eqnarray}
\label{pegado1}
 \left[ zb(T) \right]_{\Sigma} &=&\rho(\tau)= \left[R(t,r)\right]_{\Sigma} \,, \\
\label{pegado2}
      \left[ {\rm d} T^2-\left(\frac{b^2}{1+z^2}\right){\rm d}z^2\right]_{\Sigma}
      &=&{\rm d}\tau^2\ 
      = \left[ {\rm e}^{A(t,r)}{\rm d}t^2-{\rm e}^{B(t,r)}{\rm d}r^2 \right]_{\Sigma} \,,
 \end{eqnarray}
 where the subscript $\Sigma$ indicates that the metric functions of the inner and outer regions are evaluated on the shell. The above equations have the physical interpretation that observers in ${\cal M}^+$ and ${\cal M}^-$ must measure the same value for the physical radius of the bubble. The inner and outer radial coordinates of the bubble are denoted by $\zeta(T)\equiv z|_{\Sigma}$ and $x(t)\equiv r|_{\Sigma}$, respectively. Then, from Eqs.~(\ref{pegado1}) and (\ref{pegado2}) we can express $(T,\zeta(T))$ and $(t,x(t))$ as functions of $\tau$. However, since Lema\^{\i}tre's solution is known only numerically, it is convenient to describe the evolution of the shell in terms of the outer coordinates $(t,r,\theta,\phi)$. Hence, the evolution of $\Sigma$ will be parametrised by $t$, instead of $\tau$. 

 The restrictions (\ref{pegado1}) and (\ref{pegado2}), together with an appropriate jump analysis of Einstein's equations in the vicinity of the hypersurface of discontinuity $\Sigma$, lead to the Israel's junction conditions, given by \citep{Israel1966,Berezin1987,Sakai1994,Gron2007}:
 \begin{eqnarray}
 \label{IsraelI}
 -\frac{\sigma}{2}&=&\left[K^{\theta}_{\ \theta} \right]\,, \\
 \label{IsraelII}
 \Pi&=&\left[K^{\tau}_{\ \tau} \right] + \left[K^{\theta}_{\ \theta} \right]\,, \\
 \label{IsraelIII}
\frac{{\rm d}\sigma}{{\rm d}\tau} +\frac{2}{\rho}\frac{{\rm d}\rho}{{\rm d}\tau}(\sigma+\Pi)&=& -[T^n_ {\ \tau}]\,,
 \end{eqnarray}
 where all quantities are functions of the proper time on the shell, $\tau$, and $T^n_{\ \tau}\equiv {\rm \bf e}^{\alpha}_{\ \tau}T_{\alpha}^{\ \beta}n_{\beta}$ is the projection of the energy-momentum tensor of the inner/outer region in the direction normal to the shell surface. The extrinsic curvature tensor is defined as $K_{a b}\equiv n_{\alpha;\beta}{\rm \bf e}^{\alpha}_{\ a}{\rm \bf e}^{\beta}_{\ b}$, and the projectors over the hypersurface $\Sigma$ are 
 \begin{eqnarray}
 \label{proyector1}
{\rm \bf e}^{\alpha}_{\ \tau}&=&\left(\frac{{\rm d}{t}}{{\rm d}\tau},\frac{{\rm d}x}{{\rm d}\tau},0,0\right)\,, \\
{\rm \bf e}^{\alpha}_{\ \theta}&=&\left(0,0,1,0\right)\,, \\
\label{proyector3}
{\rm \bf e}^{\alpha}_{\ \phi}&=&\left(0,0,0,1\right)\,,
\end{eqnarray}
where $u^{\alpha}={\rm \bf e}^{\alpha}_{\ \tau}$ is the velocity of the bubble, and the normal vector oriented to the outer region is defined by the conditions $u^{\alpha}n_{\alpha}=0$ and $n^{\alpha}n_{\alpha}=-1$. Equation (\ref{IsraelII}) can be substituted by an equation of state for the matter content on the shell, which is assumed of the form  $\Pi=w\sigma$.

Equation (\ref{IsraelIII}) shows the energy-momentum balance in the bubble's wall, and completely determines the evolution of the shell. Note also that it is independent of the value of the inner/outer cosmological constants, since the contribution $[\Lambda g_{\mu\nu}n^\mu u^\nu]$ vanishes. Consequently, as the bubble expands there is not vacuum energy transferred from ${\cal M}^\pm$ regions to the surface energy of the bubble, and the liberated vacuum energy is completely transformed into kinetic energy of the shell. The effect of the stress-energy tensor $S_{\mu\nu}$ of the bubble on the spacetime geometry can be analysed integrating Eqs.~(\ref{EinsteinEqns}) across $\Sigma$, which yields 
\begin{eqnarray}
\label{S1}
&&S_{\mu\nu}n^\mu n^\nu = S_{\alpha\nu}n^\nu h^\alpha_\mu=0\,, \\
\label{S2}
&&S_{\mu\nu}=S_{\alpha\beta}h^\alpha_\mu h^\beta_\nu= [K_{\mu\nu}]-h_{\mu\nu}[K]\,. 
\end{eqnarray}
The last set of equations (\ref{S2}) (namely Lanczos equations) links the discontinuity in the extrinsic curvature across the shell to its energy-momentum content, while equations (\ref{S1}) have the physical meaning that no momentum associated with the surface layer flows out of $\Sigma$ (so that $S_{\mu\nu}$ lives on the hypersurface of the shell and is well defined by Eq.~(\ref{Sintegral})) \cite{Laguna-Castillo1986}. 

 The history of the shell is then completely determined by  Eqs.~(\ref{IsraelI}) and (\ref{IsraelIII}), which we need to rewrite in terms of the external coordinates $t$ and $r$ for the numerical computation. We will follow \citep{Fischler2008} to get the appropriate evolution equations.

Let us start calculating the  angular components of the extrinsic curvature tensor for each region, $K_{\theta}^{\theta}=h^{\theta\theta}n_{\theta;\theta}$, which become
\begin{eqnarray}
(K_{\theta}^{\theta})^-&=&  - \frac{\gamma_{-}}{\rho\sqrt{1+\zeta^2}} \left(\zeta b \frac{{\rm d}b}{{\rm d}T} \frac{{\rm d} \zeta}{{\rm d}\tau} + (1+\zeta^2)\frac{{\rm d}T}{{\rm d}\tau}\right) \ \ \,, \\
(K_{\theta}^{\theta})^+ & =& -\frac{\gamma_+}{\rho {\rm e}^{A/2}{\rm e}^{B/2}}\left({\rm e}^A R'\frac{{\rm d}t}{{\rm d}\tau} +{\rm e}^B \dot{R}\frac{{\rm d}x}{{\rm d}\tau}  \right)\,,
\end{eqnarray}
where all the metric functions are evaluated at $\Sigma$, and $\gamma_{\pm}=1$ ($\gamma_{\pm}=-1$) if the shell is expanding (collapsing). The explicit form of $K^{\theta}_{\theta}$ allows us  to express the first derivative of restriction (\ref{pegado1}) as 
 \begin{equation}
\label{dotrho2}
\left(\frac{{\rm d}\rho}{{\rm d}\tau}\right)^2=\Delta^\pm + [\rho(K_{\theta}^{\theta})^{\pm}]^2 \,,
\end{equation} 
with
\begin{eqnarray}
\label{Delta1}
\Delta^+&=&-1+\left(\frac{2M}{R^3}+\frac{\Lambda^+}{3}\right)\rho^2\,,  \\
\label{Delta2}
 \Delta^-&=&-1+\frac{\Lambda^-}{3}\rho^2\,.
\end{eqnarray}
  Equation (\ref{IsraelI}) can be now rewritten as
\begin{equation}
\label{Fischler1}
-\gamma_+\sqrt{\left(\frac{{\rm d}\rho}{{\rm d}\tau}\right)^2-\Delta^+}+\gamma_-\sqrt{\left(\frac{{\rm d}\rho}{{\rm d}\tau}\right)^2-\Delta^-}=-\frac{\sigma \rho}{2}\,,
\end{equation}
where the arguments of the square roots are always positive due to (\ref{dotrho2}). After some algebra, and replacing expressions (\ref{Delta1}) and (\ref{Delta2}), we get 
\begin{equation}
\label{dotrho1}
\left(\frac{{\rm d}\rho}{{\rm d}\tau}\right)^2=\rho^2V^2-1\,,
\end{equation}
 where  
 \begin{equation}
 \label{def_V}
 V^2\equiv\frac{\Lambda^-}{3}+\left[\frac{\sigma}{4}+\frac{1}{\sigma}\left(\frac{\Lambda^+-\Lambda^-}{3}+\frac{2M}{R^3}\right) \right]^2\,.
 \end{equation}
Since  $({\rm d}\rho/{\rm d}\tau)=[\dot{R}+R'({\rm d}x/{\rm d}t)]({\rm d}t/{\rm d}\tau)$ on the shell, Eq.~(\ref{dotrho1})
  yields a quadratic equation for $({\rm d}x/{\rm d}t)$,
  which solutions are given by 
 \begin{equation}
 \label{dxdt_gral}
 \frac{{\rm d}x}{{\rm d}{t}} = \frac{-\dot{R}R'\pm\sqrt{(R^2V^2-1)[R'^2{\rm e}^A-\dot{R}^2{\rm e}^B+{\rm e}^A{\rm e}^B(R^2V^2-1)]}}{R'^2+{\rm e}^B(R^2V^2-1)}\,.
 \end{equation}
Since we are interested in solutions $x(t)$ such that 
$({\rm d}x/{\rm d}t)$ is initially positive (expanding bubbles for the initial conditions given in Section~\ref{sec:burbuja_num}), we choose the positive sign for the numerical integration. A restriction on the function $x(t)$ for the motion of the bubble follows from imposing that effectively the r.h.s of Eq.~(\ref{dotrho1}) be positive on the shell, which leads to 
\begin{equation}
\label{cotax}
 1<R^2V^2\,. 
 \end{equation}
Note that if the evolution reaches values of $x(t)$ such that $1=V^2R^2$, then the proper velocity of the bubble vanishes. 
 
 It only remains to rewrite Eq.~(\ref{IsraelIII}) in the outer coordinates. The outer projection of the energy-momentum tensor normal to $\Sigma$ is 
\begin{equation}
(T^n_{\tau})^+ =-\gamma_+\frac{{\rm d}t}{{\rm d}\tau}\frac{{\rm d}x}{{\rm d}\tau}\frac{{\rm e}^{A(t,x)/2}{\rm e}^{B(t,x)/2}[\epsilon(t,x)+p(t,x)]}{\sqrt{{\rm e}^{A(t,x)}-{\rm e}^{B(t,x)}\left(\frac{{\rm d}x}{{\rm d}t}\right)^2}}\,,
\end{equation}
and, since the bubble encloses a vacuum region, we have $(T^n_{\tau})^-=0$. Hence Eq.~(\ref{IsraelIII}) takes the form
 \begin{equation}
 \label{dsigdt_gral}
 \frac{{\rm d}\sigma}{{\rm d}{t}} = -2(1+w)\sigma\frac{\dot{R}}{R}  +\gamma_+(\epsilon + p)\frac{{\rm d}x}{{\rm d}{t}}\frac{{\rm e}^{A/2}{\rm e}^{B/2}}{\sqrt{{\rm e}^{A}-{\rm e}^{B}\left(\frac{{\rm d}x}{{\rm d}{t}}\right)^2}}\,.
 \end{equation}
 We will assume that the matter on $\Sigma$ satisfies the weak energy condition during all the evolution, that is, $\sigma>0$. 
 This condition is equivalent to impose the following restrictions\footnote{In a general spherically-symmetric case, described by the line element (\ref{ds_Lem}), the auxiliary quantity $\Delta$ is defined as $\Delta\equiv(\dot{R}^2/{\rm e}^{A})-(R'^2/{\rm e}^{B})$.}  
 \begin{eqnarray}
 \label{cond_sigma}
 \Delta^+ -\Delta^-&>& \frac{\rho^2 \sigma^2}{4}\,,\ {\rm if}\ \gamma=+1\,, \\
 \label{cond_sigma2}
\Delta^+ -\Delta^-&<& \frac{\rho^2 \sigma^2}{4}\,,\ {\rm if}\ \gamma=-1\,.
\end{eqnarray}
 
The outer geometry and the coupled system given by Eqs.~(\ref{dxdt_gral}) and (\ref{dsigdt_gral}) determine the evolution of the shell in terms of the external coordinates $(t,x(t))$, which must be calculated 
through numerical integration. 
%
 %
 %
 \subsection{Numerical evolution}
 \label{sec:burbuja_num}
  We have developed a numerical code to compute the evolution of the bubble, given by the solution of equations (\ref{dxdt_gral}) and (\ref{dsigdt_gral}).  These equations  are coupled to those determining the evolution of the external geometry, which can be written as follows 
  \citep{Alfedeel2010}
 \begin{eqnarray}
 \label{dRdt_Lem}
     \dot{R}&=&{\rm e}^{A/2}\left[\frac{2M}{R}+\frac{\Lambda^+}{3}R^2-1 +R'^2{\rm e}^{-B}\right]^{1/2}\,,  \\
     \label{dMdt_Lem}
      \dot{M}&=&-\frac{p}{2}R^2\dot{R}\,,  \\
      \label{depsilondt_Lem}
     \dot{\epsilon}&=&-p'\frac{\dot{R}}{R'}-[\epsilon+p]\left[\frac{\dot{R}'}{R'}+2\frac{\dot{R}}{R}\right] \,,  \\
     \label{dBdt_Lem}
     \dot{B}&=&2\left[\frac{\dot{R}'}{R'}+\frac{\dot{R}p'}{[\epsilon+p]R'}\right]\,, 
     \end{eqnarray}
     with
     \begin{eqnarray}
     \label{def_drM}
     M'&=& \frac{\epsilon}{2} R^2 R'\,, \\
      \label{def_Atr} 
    A&=&-2\int_{0}^r\frac{p'}{\epsilon+p}{\rm d}r\,.
  \end{eqnarray}
  The external pressure $p$ is determined from the corresponding equation of state for the outer matter content. 
  
 The integration of the above system of partial differential equations was implemented using the  method of lines with a fourth order differentiation scheme \citep{Schiesser1991}. 
  We choose for our problem the following initial profiles:
\begin{eqnarray}
&& R(t_0,r)=a_0r\,, \\
&& E(r)=-\frac{1}{2}kr^2\,,  \\ 
&&\epsilon(t_0,r)=\epsilon_0 \left[1-\delta_{\epsilon}{\rm exp}\left(-\frac{(r-r_0)^2}{s_0^2}\right)\right]\,,
\end{eqnarray}
which are sufficient to completely determine the evolution of the outer geometry.\footnote{The functions $M(t_0,r)$ and $A(t_0,r)$ are then computed from Eqs.~(\ref{def_drM}) and (\ref{def_Atr}), respectively, and ${\rm e}^{B(t_0,r)}=R'^2(t_0,r)/(1+2E(r))$. It is also possible to introduce the inhomogeneous profile through the curvature function $E(r)$, as discussed in \citep{Bolejko2006,Fischler2008}.} In particular, we consider $a_0=1$ in order to initially set the radial coordinate of the bubble equal to its proper radius.
The curvature is characterised by the constant $k=(\Lambda^-/3)/10$, which is low enough to ensure that the evolution given by equation (\ref{dRdt_Lem}) is initially dominated by its two first terms.\footnote{ This condition is a first attempt to analyse the evolution of the shell in inhomogeneous radiation backgrounds and can be relaxed in an extended 
 analysis of the present work. } 
 The quantities $\delta_{\epsilon}$, $r_0$ and $s_0$ determine the inhomogeneous initial distribution of the background matter, and the constant $\epsilon_0$ represents the asymptotic value of the background density away from the inhomogeneous region. 

The parameter $\Lambda^-$ represents the vacuum energy of the region inside the bubble, and is intrinsically related to the energy scale imposed by the inflationary models for the nucleation process \citep{Guth1981}. 
We choose $\Lambda^-\simeq 5 \times 10^{-5}$  to characterise the inner region, which corresponds to an energy of order $~10^{14}\ {\rm GeV}$ in Planck units.  Since there are not {\it a priori} restrictions on the  parameters $\Lambda^+$ and $\epsilon_0$, we shall work with values $\Lambda^+<\epsilon_0=10\Lambda^-$, which ensure that at $t=t_0$ the dynamics of the external region is dominated by the term  $(2M/R)$ in Eq.~(\ref{dRdt_Lem}). Consequently, the potential effects on the dynamics due to the background dust or radiation distributions become more pronounced. In the opposite case, the $\Lambda^+$ dominated expansion would rapidly dilute the background density, thus becoming a de Sitter-de Sitter scenario. The parameter $\Lambda^+$ is allowed to take four representative values: $\Lambda^+=0,\Lambda^-/2,\Lambda^-,2\Lambda^-$.
 Each of these leads to a different dynamical behaviour, which will be analysed in the following sections.
 
Finally, the initial conditions for the thin-shell are   
 $x_0=15$ and  $\sigma_0=1 \times 10^{-3}$. 
The election of $x_0$ is such that the nucleation of the bubble takes place at a point where the gradient of $\epsilon$ is non-negligible.
 For the most general case, in which the bubble expands in an inhomogeneous background with non-zero pressure (described by Lema\^{\i}tre's solution), this choice implies also a non-zero initial pressure gradient,  whose influence on the background evolution is briefly analysed in  Appendix~\ref{appLEM}. 
On the other hand, the initial value for $\sigma_0$ is chosen to satisfy the constraint given by Eq.~(\ref{cond_sigma}). We consider the values $w=0,1/3$ for the equation of state parameter for the matter on the bubble.

The results for the numerical evolution of the vacuum bubble embedded in different backgrounds are shown in the next sections. We will start with the discussion of the simplest case (namely, homogeneous outer regions), with the aim of  studying first the effects of the radiation pressure over the bubble evolution.  Afterwards, we will focus the analysis on the features due to inhomogeneous distributions. 
%
%
 \section{Evolution in homogeneous backgrounds: dust vs. radiation}
 \label{sec:PvoRad}
   We shall study in this section the effects of two different homogeneous backgrounds on the dynamics of the bubble, corresponding to  contents of dust or radiation. In  both cases, the outer region is characterised by the isotropic and homogeneous FLRW metric, with line element given by
  \begin{equation}
     {\rm d}s^2={\rm d}t^2-a^2(t)\left(\frac{1}{1-kr^2}{\rm d}r^2-r^2{\rm d}\Omega^2\right)\,.
  \end{equation}
This metric can be recovered from the expression (\ref{ds_Lem}) when $E(r)=-\frac{1}{2}kr^2$, $R(t,r)= a(t)r$ and $\epsilon(t,r)=\epsilon(t)$. In this case we have that $p'=0$, and hence $A(t,r)=0$ and $e^{B(t,r)}=R'^2(t,r)/(1+2E(r))=a^2(t)/(1-kr^2)$. Eqs.~(\ref{dRdt_Lem})-(\ref{dBdt_Lem}), which determine the evolution of the FLRW  geometry, are simplified to the following:
 \begin{eqnarray}
 \dot{a}^2&=&\frac{2M}{ar^3}+\frac{a^2\Lambda^+}{3}-k\,, \\
  \dot{M}&=&-\dot{a}a^2r^3\frac{p}{2}\,,\\
 \dot{\epsilon}&=&-3(\epsilon+p)\frac{\dot{a}}{a}\,,
 \end{eqnarray}
 along with the equations of state $p=0$ (dust) or $p=\epsilon/3$ (radiation). In the case of a dust background, we also have $\dot{M}(t,r)\equiv 0$.
 Equations (\ref{dxdt_gral}) and (\ref{dsigdt_gral}), which respectively determine the evolution of the radial coordinate and the energy density of the bubble, become
 \begin{eqnarray}
\label{dxdt_FLRW}
\frac{{\rm d}x}{{\rm d}{t}}&=&\frac{-(1-kx^2)\dot{a}+\sqrt{(x^2a^2V^2-1)(1-kx^2)(a^2V^2-\dot{a}^2-k)}}{xa(a^2V^2-k)}\,, \\
 \label{dsigdt_FLRW}
  \frac{{\rm d}\sigma}{{\rm d}{t}}&=&-2(\sigma+\Pi)\frac{\dot{a}}{a}+\gamma_+\frac{{\rm d}x}{{\rm d}{t}}
 \frac{a(\epsilon + p)}{\sqrt{1-kx^2}}\frac{1}{\sqrt{1-\frac{a^2}{1-kx^2}\left(\frac{{\rm d}x}{{\rm d}{t}}\right)^2}}\,,
 \end{eqnarray}
 with
 \begin{equation}
  V^2=\frac{\Lambda^-}{3}+\left[\frac{\sigma}{4}+\frac{1}{\sigma}\left(\frac{\Lambda^+-\Lambda^-}{3}+\frac{2M}{a^3x^3}\right) \right]^2\,.
 \end{equation}
 
\begin{figure}[tbp]
\centering 
  \includegraphics[angle=-90,width=7.5cm]{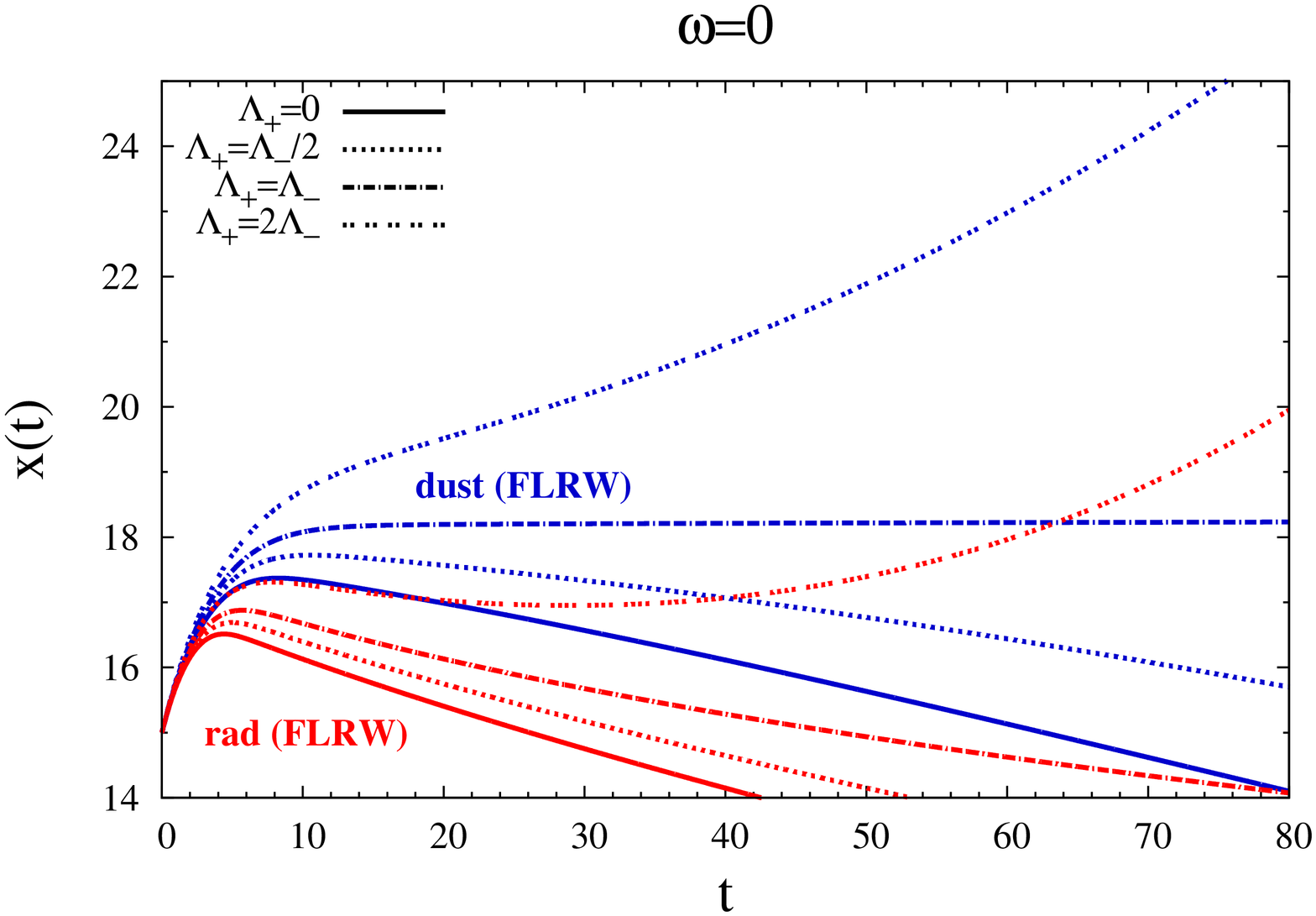}
  \hfill
  \includegraphics[angle=-90,width=7.5cm]{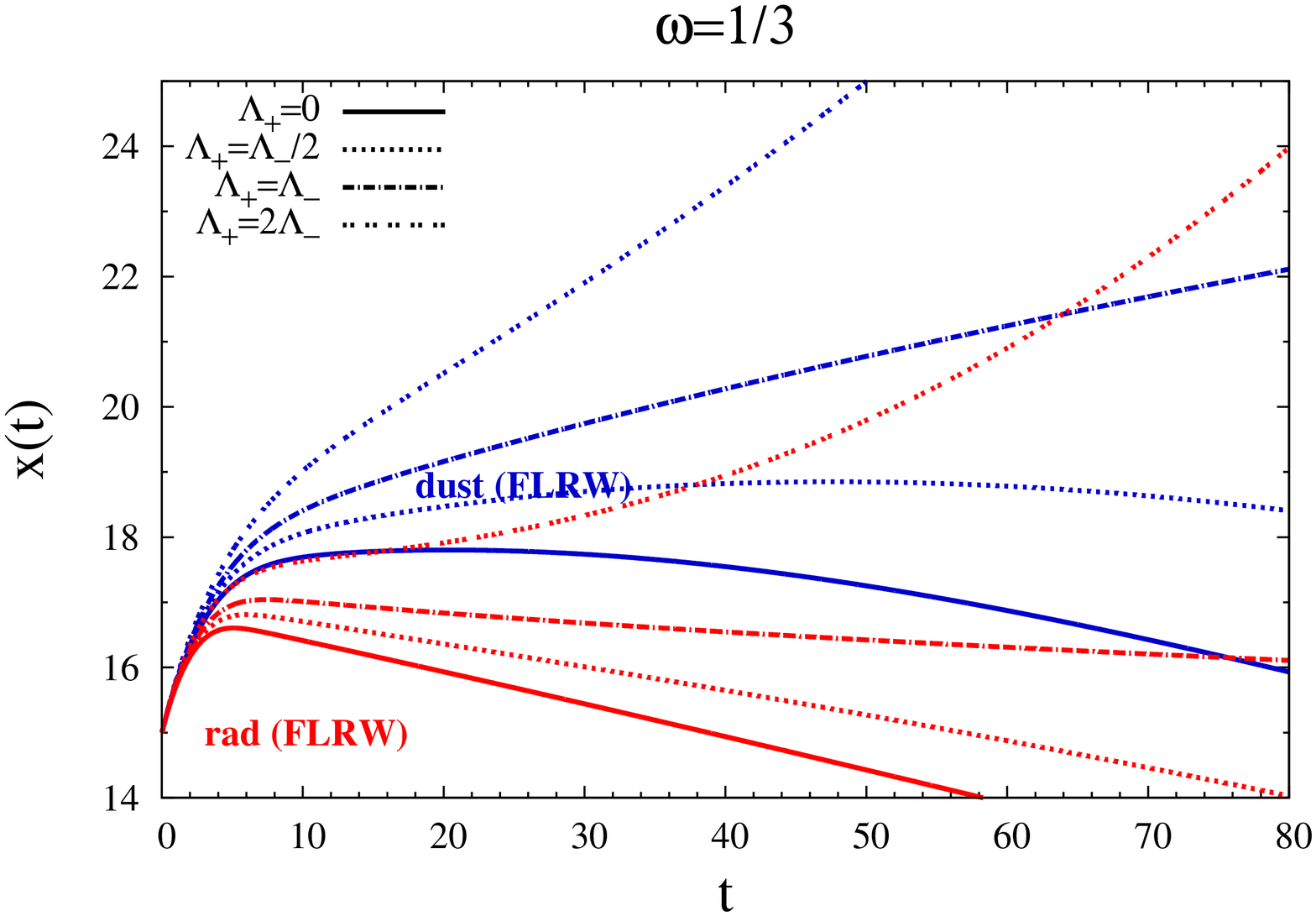}
\caption{Evolution of the external radial coordinate of the bubble for different values of the parameter $\Lambda^+$. The left and right panels correspond, respectively, to $w=0$ and $w=1/3$, where $w$ is  the parameter of the equation of state of the bubble. We choose $\delta_{\epsilon}=0$ to describe the homogeneous background. If $\Lambda^+<\Lambda^-$, the radius expressed in external coordinates initially grows but then decreases, whereas for $\Lambda^-<\Lambda^+$ cases it grows monotonically. The parameters which characterised the geometry, as well as the initial conditions, are those detailed in section~\ref{sec:burbuja_num}. } 
\label{fig:PvoRad}
\end{figure}

\begin{figure}[tbp]
\centering
  \includegraphics[angle=-90,width=7.5cm]{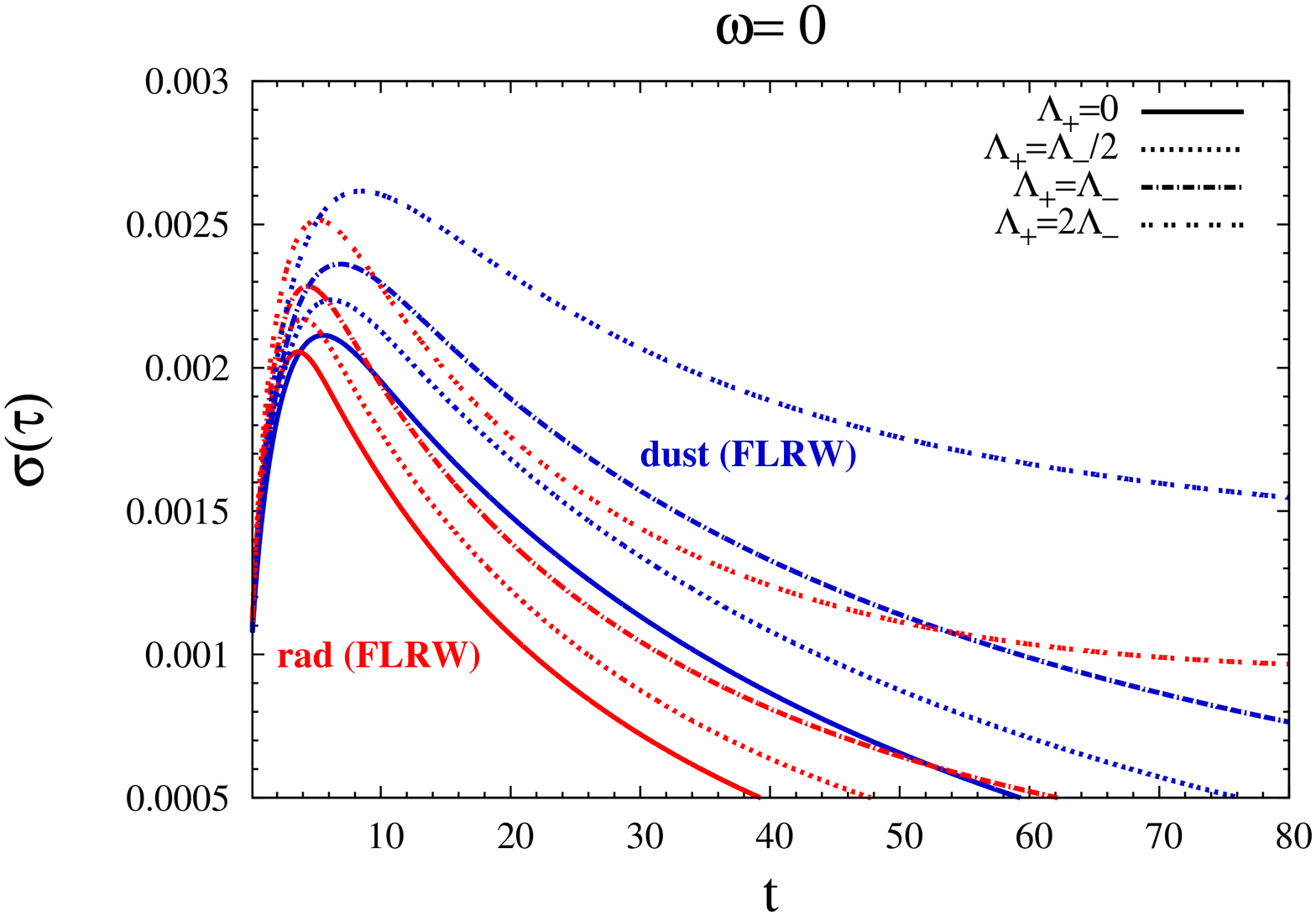}
  \hfill
  \includegraphics[angle=-90,width=7.5cm]{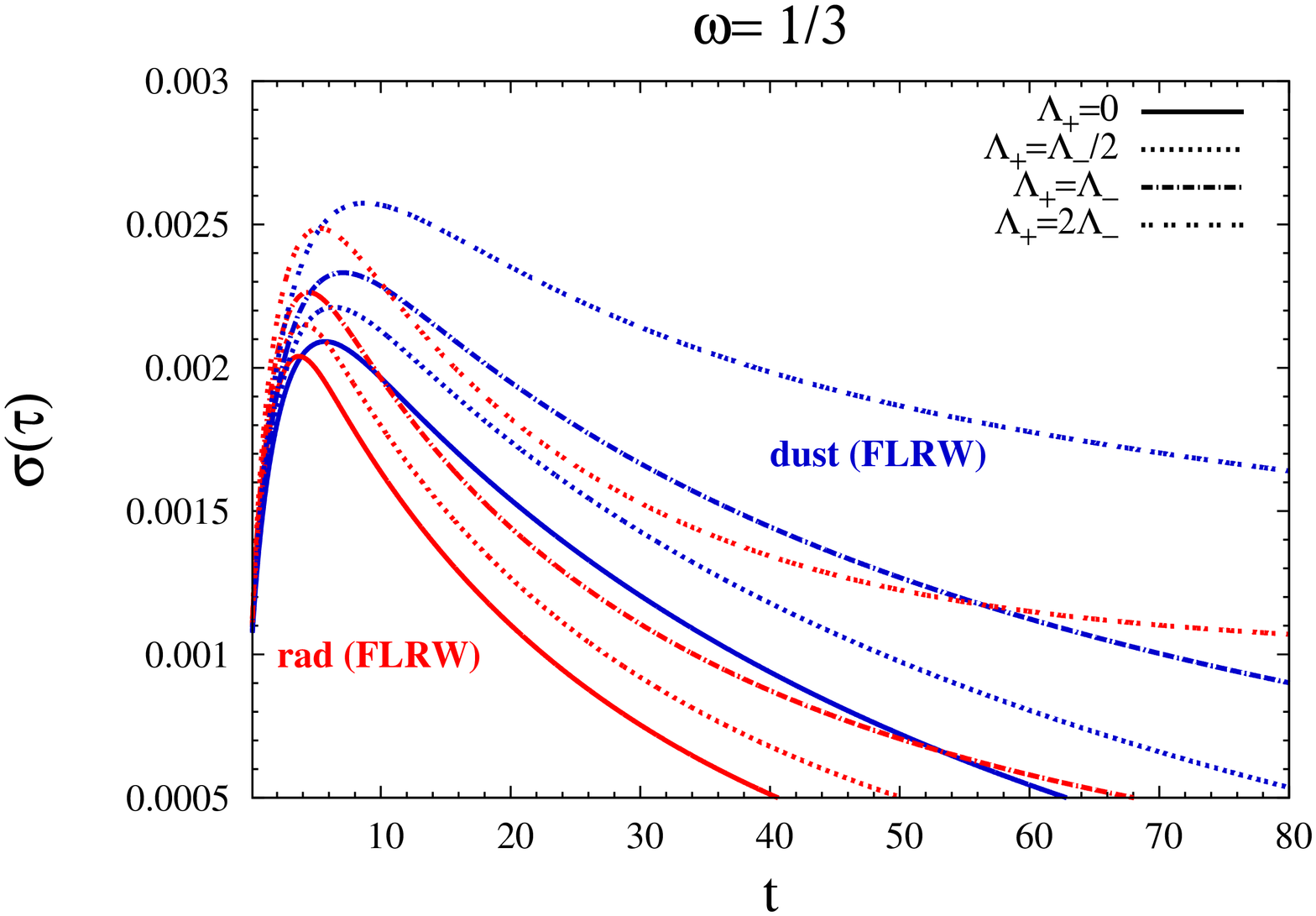}
\caption{Evolution of the energy density of the thin-shell with  $w=0$ (left) and $w=1/3$ (right). For all values of $\Lambda^+$, the evolution in radiation backgrounds reaches lower energy densities than the corresponding dust cases.} 
\label{fig:PvoRad_sigma}
\end{figure}

  We shall compare next the evolution of $x(t)$ and $\sigma(t)$ in figures~\ref{fig:PvoRad} and \ref{fig:PvoRad_sigma} for different homogeneous cases, and for times such that $\Lambda_+$ does not dominate the evolution. 
       The curves in figure~\ref{fig:PvoRad} show that the evolution of the radial coordinate of the bubble in the radiation background is slower than that in the corresponding dust case. In other words, for all the examples considered with the same initial conditions, the radiation background slows down the evolution of the shell. The plots also indicate that the value $w=0$ yields a slower evolution than the case with $w=1/3$. This feature can be understood as a consequence of the pressure generated by the matter content of the bubble. 
       Note however that the evolution is qualitatively the same in both cases.
       
       Noticeable differences exist between evolutions with different values of the parameter $\Lambda^+$: whereas the radial coordinate of the shell indefinitely grows if $\Lambda^-<\Lambda^+$, it eventually decreases in those cases for which $\Lambda^+<\Lambda^-$ until reaching the lower limit imposed by the constraint (\ref{cotax}), that is $x>1/(aV)$. Those cases with $\Lambda^-<\Lambda^+$, and for times large enough such that the matter density of the background is diluted, evolve asymptotically to that of de Sitter, which can be obtained in a closed form \citep{Fischler2008, Ng2011} and constitute a test for our numerical computation. 
       The corresponding evolution of the energy density of the thin-shell is shown in figure~\ref{fig:PvoRad_sigma}. In both $w=0$ and $w=1/3$ cases, $\sigma(t)$ displays lower values for the evolution in the radiation background, in agreement with the above-mentioned differences found in the evolution of the radial coordinate $x(t)$. 
       

\section{Evolution of vacuum bubbles in inhomogeneous backgrounds }
\label{sec:inhom_back}
 With the aim of analysing the effects of inhomogeneous distribution of the outer matter on the evolution of the shell, we will focus in this section on the evolution of vacuum bubbles embedded in backgrounds characterised by inhomogeneous dust or radiation distributions.
 The evolution of vacuum bubbles in inhomogeneous pressureless backgrounds, described by the spherically-symmetric LTB solution, has been previously studied in \citep{Fischler2008,Simon2009,Rakic2010}. We start by presenting these cases, but using different initial conditions and inhomogeneous profiles.  
Afterwards we will study the evolution of vacuum bubbles in inhomogeneous radiation backgrounds, described by Lema\^{\i}tre's solution. This problem has not been previously studied and represents the most important contribution of the present work.

 \subsection{Inhomogeneous dust backgrounds described by LTB geometry}
 \label{sec:LTBvsPvo}

An inhomogeneous spherically-symmetric dust background is described by the LTB metric with line element  
  \begin{equation}
{\rm d}s^2= {\rm d}t^2- \frac{R'^2(t,r)}{1+2E(r)} {\rm d}r^2 -R^2(t,r){\rm d}\Omega^2\,,
\end{equation}
which is a special case of Eq.~(\ref{ds_Lem}) with  $p(t,r)\equiv 0$. The equations that determine the evolution of the outer geometry then become
   \begin{eqnarray}
 \label{dRdt_LTB}
     \dot{R}^2&=&\frac{2M}{R}+\frac{\Lambda^+}{3}R^2+2E(r)\,,  \\
    \label{depsilondt_LTB}
     \dot{\epsilon}&=&-\epsilon\left(\frac{\dot{R}'}{R'}+2\frac{\dot{R}}{R}\right) \,,
     \end{eqnarray}
  and $M$ is a function of $r$ only. The radial coordinate and the energy density of the bubble evolve following 
\begin{eqnarray}
 \label{dxdt_LTB}
 \frac{{\rm d}x}{{\rm d}{t}}&=& \frac{-(1+2E)\dot{R}+\sqrt{(R^2V^2-1)(1+2E)(2E-\dot{R}^2+R^2V^2)}}{R'(2E+R^2V^2)}\,, \qquad\\
 \label{dsigdt_LTB}
 \frac{{\rm d}\sigma}{{\rm d}{t}}&=&-\frac{2(\sigma+\Pi)\dot{R}}{R}  +\gamma_+ \frac{{\rm d}x}{{\rm d}{t}}\frac{R'}{\sqrt{1+2E}}\frac{(\epsilon + p)}{\sqrt{1-\frac{R'^2}{(1+2E)}\left(\frac{{\rm d}x}{{\rm d}{t}}\right)^2}}\,, 
 \end{eqnarray}
with
 \begin{equation}
 V^2=\frac{\Lambda^-}{3}+\left[\frac{\sigma}{4}+\frac{1}{\sigma}\left(\frac{\Lambda^+-\Lambda^-}{3}+\frac{2M}{R^3}\right) \right]^2\,.
 \end{equation}

\begin{figure}[tbp]
\centering
 \includegraphics[angle=-90,width=7.5cm]{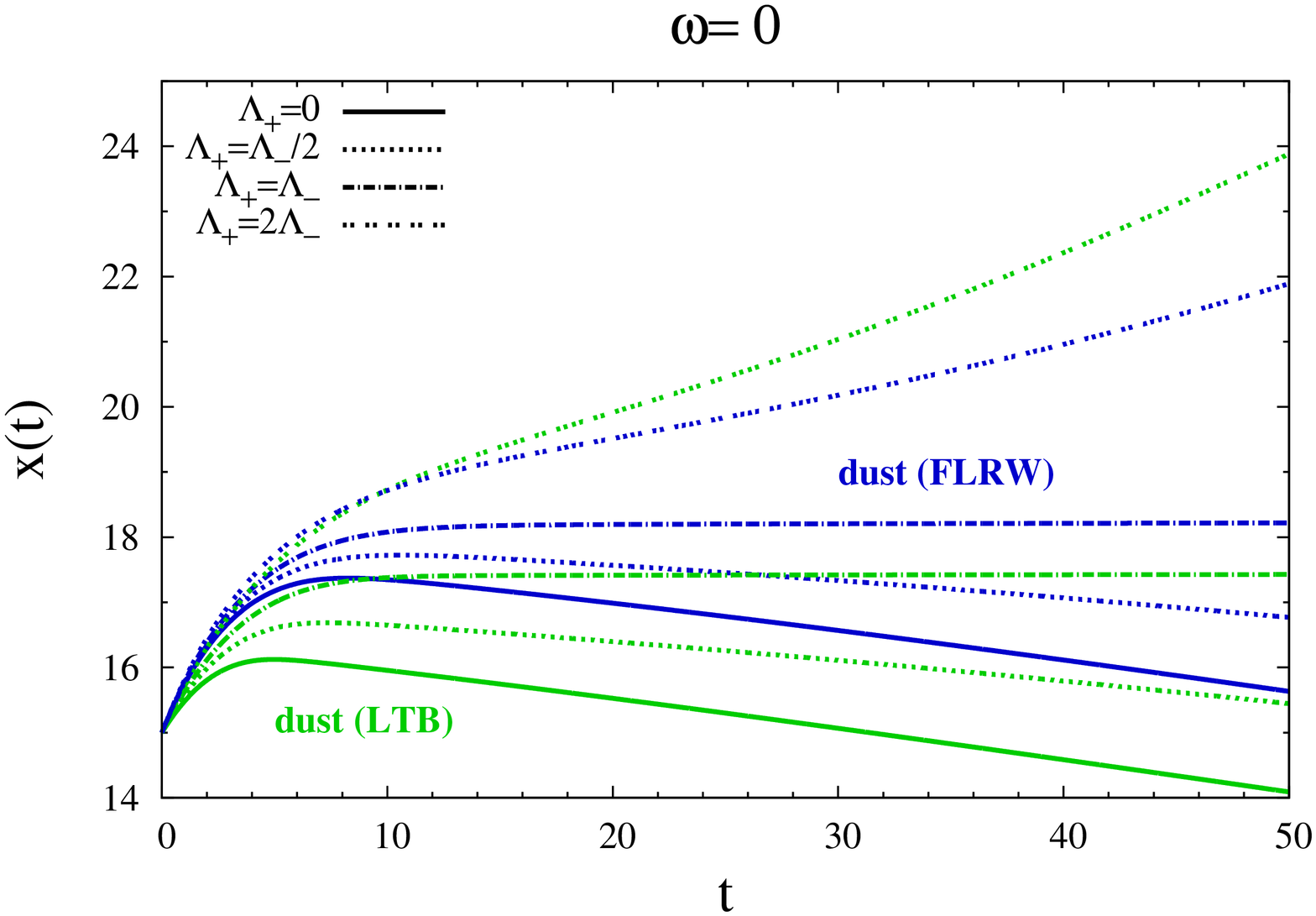}
 \hfill
 \includegraphics[angle=-90,width=7.5cm]{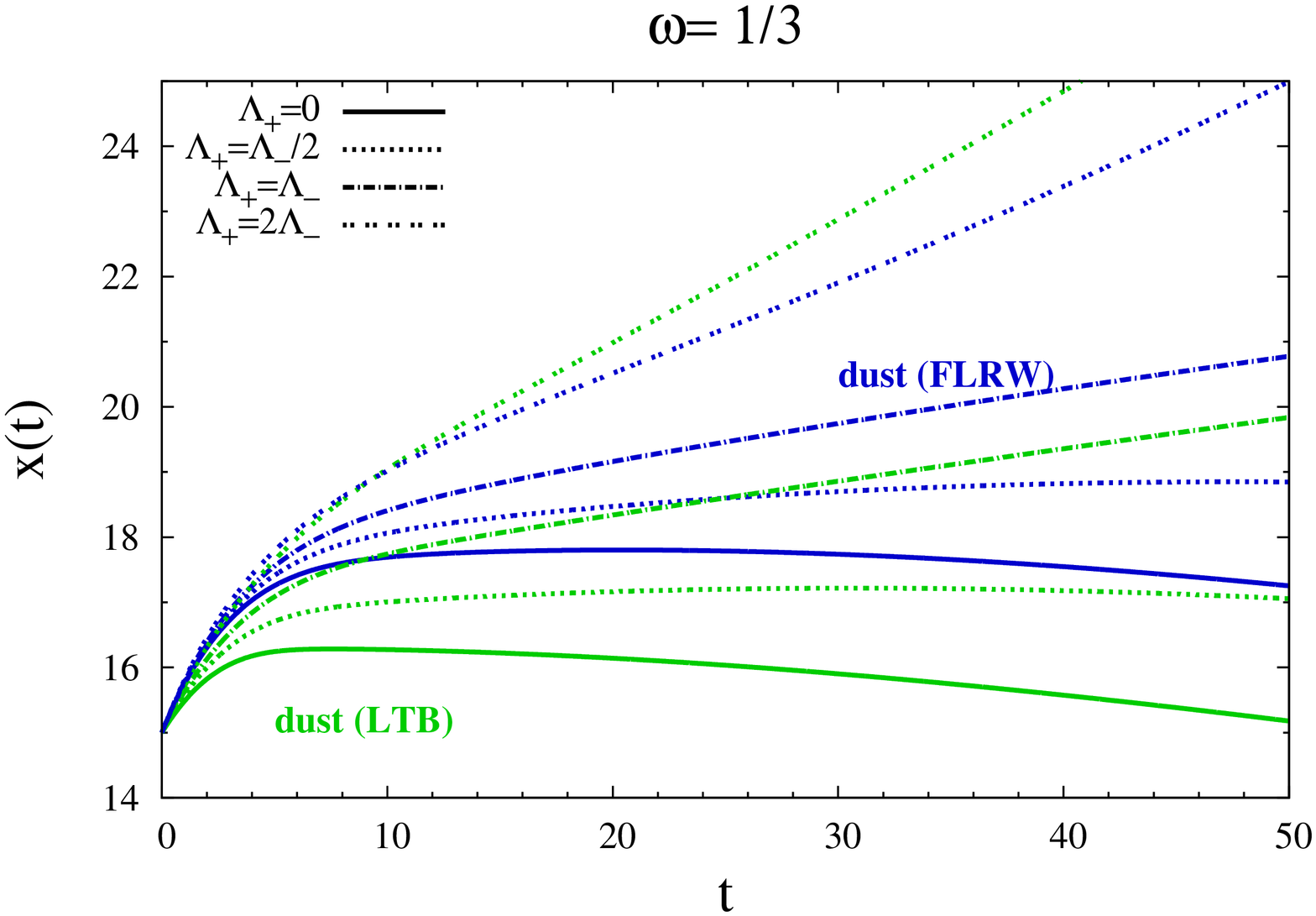}
  \caption{Evolution of the radial coordinate of the bubble immersed in an inhomogeneous background described by the LTB solution. We choose for our examples $r_0=20$, $s_0=5$ and $\delta_{\epsilon}=0.5$ ($\delta_{\epsilon}=0$ in the homogeneous case).
  The curves are compared with those obtained for homogeneous dust backgrounds with density initially equal to $\epsilon_0$.}
\label{fig:PvoLTB}
\end{figure}

 The evolution of the external radial coordinate of the bubble in an inhomogeneous dust background, considering $w=0,1/3$ and different values of the parameter $\Lambda^+$, is shown in figure~\ref{fig:PvoLTB}. In order to analyse the effects produced by the outer inhomogeneities on the evolution of the bubble, we compare the curves with those obtained for homogeneous dust backgrounds with initial homogeneous density equal to the asymptotic value $\epsilon_0$. We can observe that the growth of the radial coordinate in inhomogeneous backgrounds is slower than the corresponding homogeneous evolutions when the bubble is initially located in a sub-density region. Although the expansion of the background will dilute the inhomogeneous external region, and then the radial coordinate will eventually follow an homogeneous evolution, it is important to highlight that inhomogeneous profiles yield evolutions that are quantitatively different of those in the homogeneous case.

\subsection{Inhomogeneous radiation backgrounds described by Lema\^{\'i}tre's geometry}
  We have studied in the previous sections possible effects on the evolution of vacuum bubbles due to (i) the pressure of homogeneous radiation backgrounds (Sect.~\ref{sec:PvoRad}), and (ii) the nucleation of bubbles in sub-density regions of inhomogeneous dust environments (Sect.~\ref{sec:LTBvsPvo}).
  Motivated by these analysis, we will focus in this section on exploring the problem which combines both effects, that is, the evolution of vacuum bubbles in  inhomogeneous radiation backgrounds. The external geometry is described in this case by Lema\^{\i}tre's solution, while the radial coordinate and the energy density of the bubble obey Eqs.~(\ref{dxdt_gral}) and (\ref{dsigdt_gral}).
  
\begin{figure}[tbp]
\centering 
\includegraphics[angle=-90,width=7.5cm]{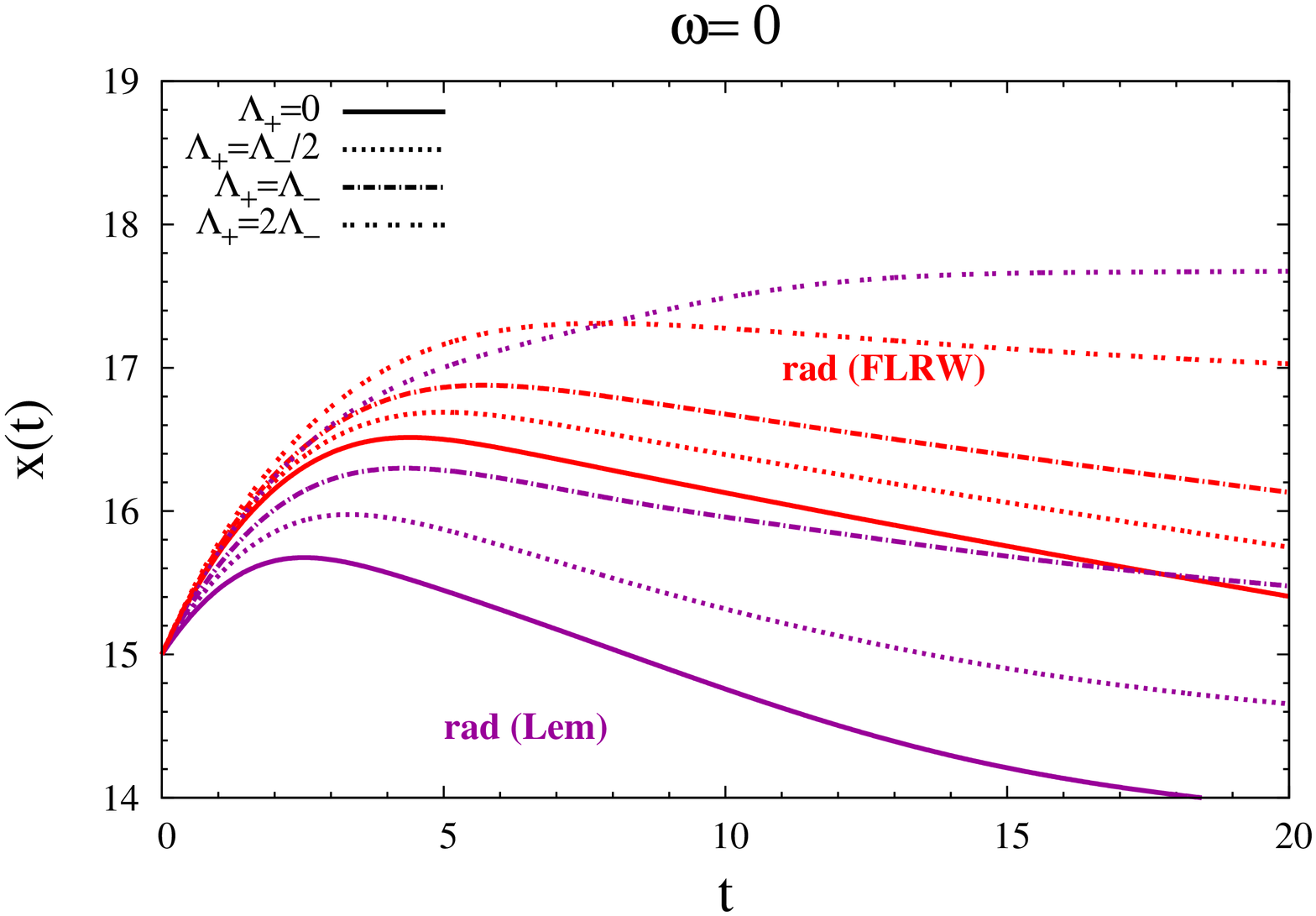}
\hfill
\includegraphics[angle=-90,width=7.5cm]{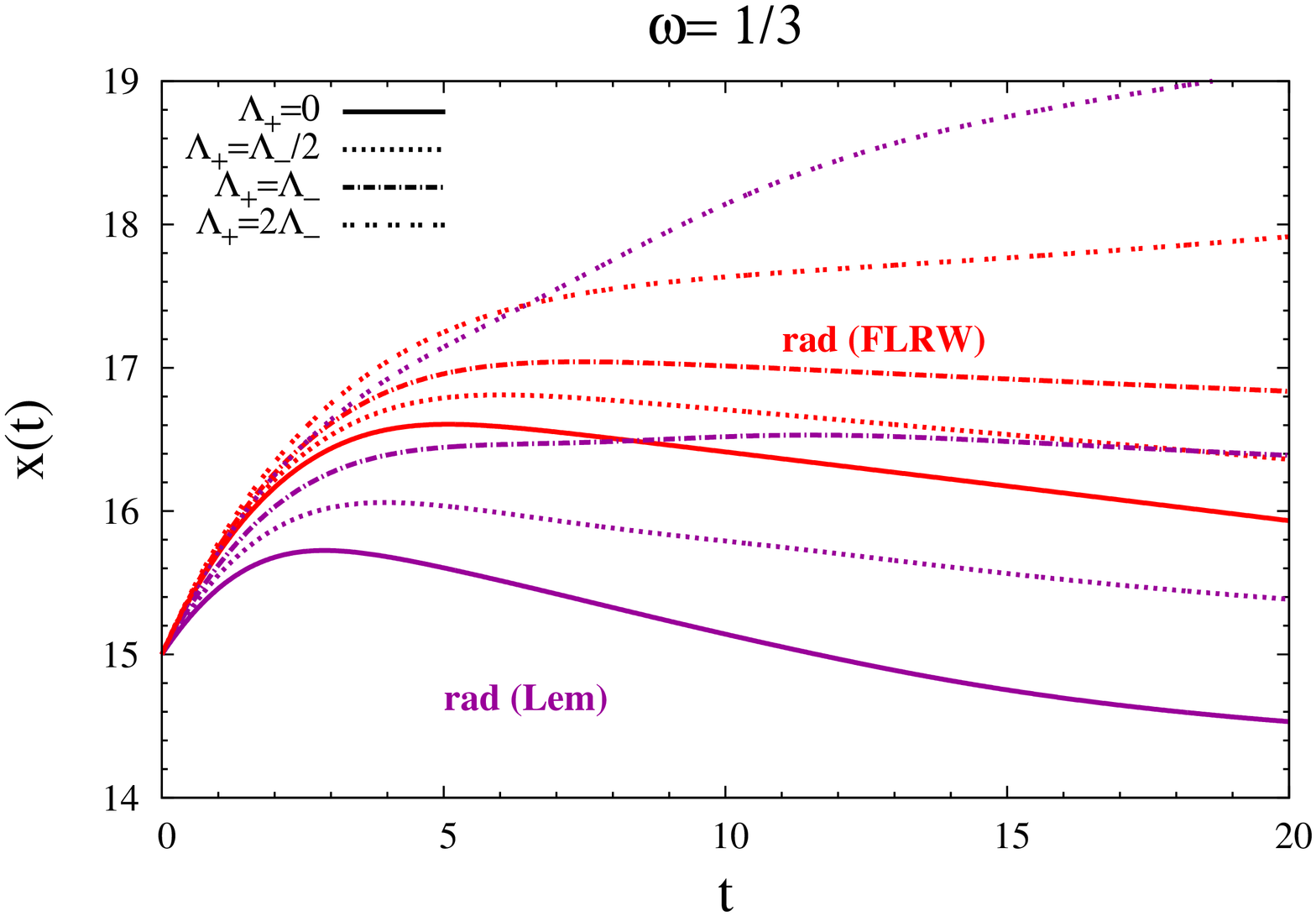}
\caption{Evolution of the radius of the bubble in the case of inhomogeneous radiation backgrounds described by the Lema\^{\i}tre's solution.  We choose for our examples $r_0=20$, $s_0=5$ and $\delta_{\epsilon}=0.5$ ($\delta_{\epsilon}=0$ in the homogeneous case). The FLRW curves represent the evolution in homogeneous radiation backgrounds with density initially equal to $\epsilon_0$.}
\label{fig:LemVsrad}
\end{figure}

\begin{figure}[tbp]
\centering 
  \includegraphics[angle=-90,width=7.5cm]{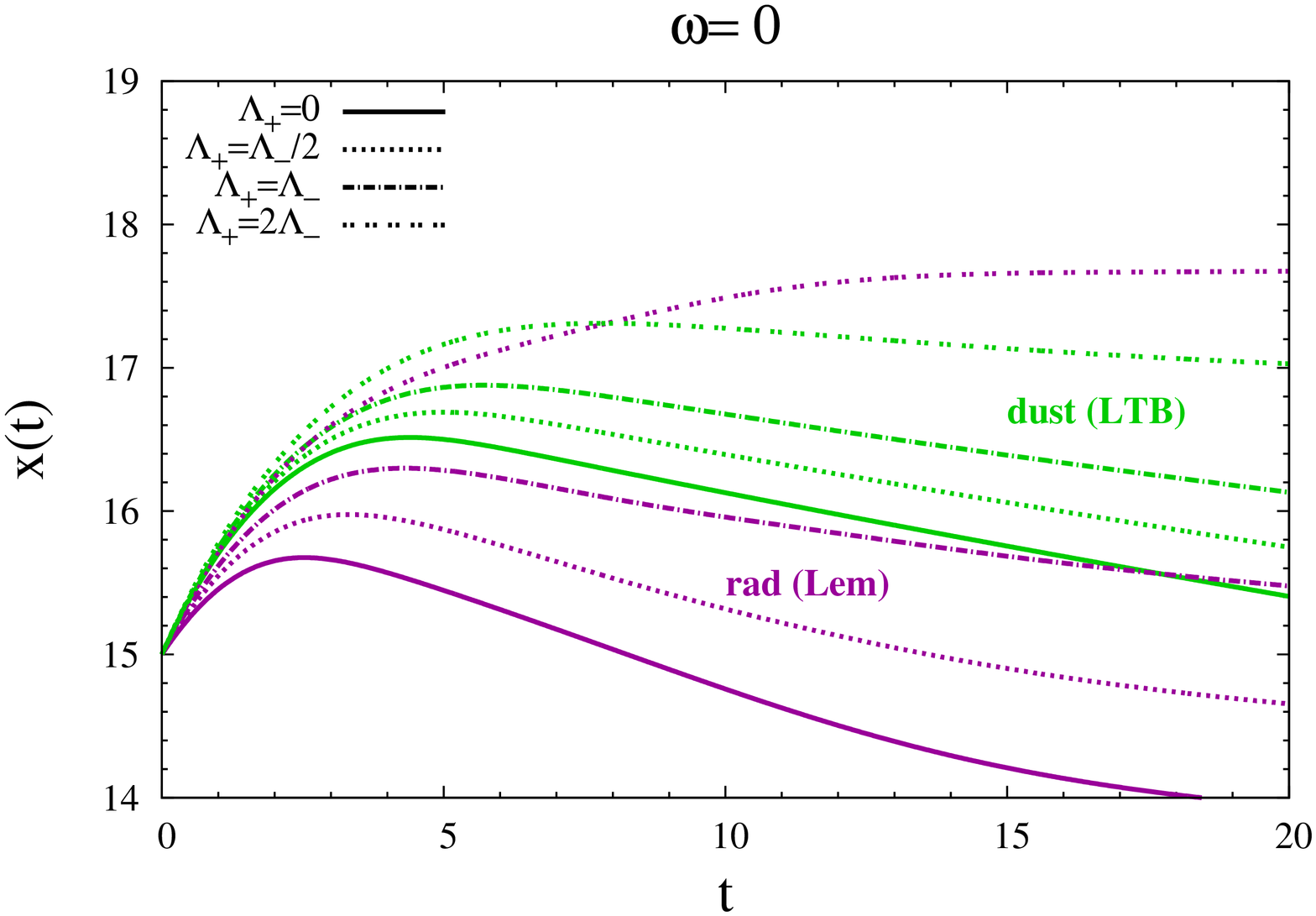}
  \hfill
  \includegraphics[angle=-90,width=7.5cm]{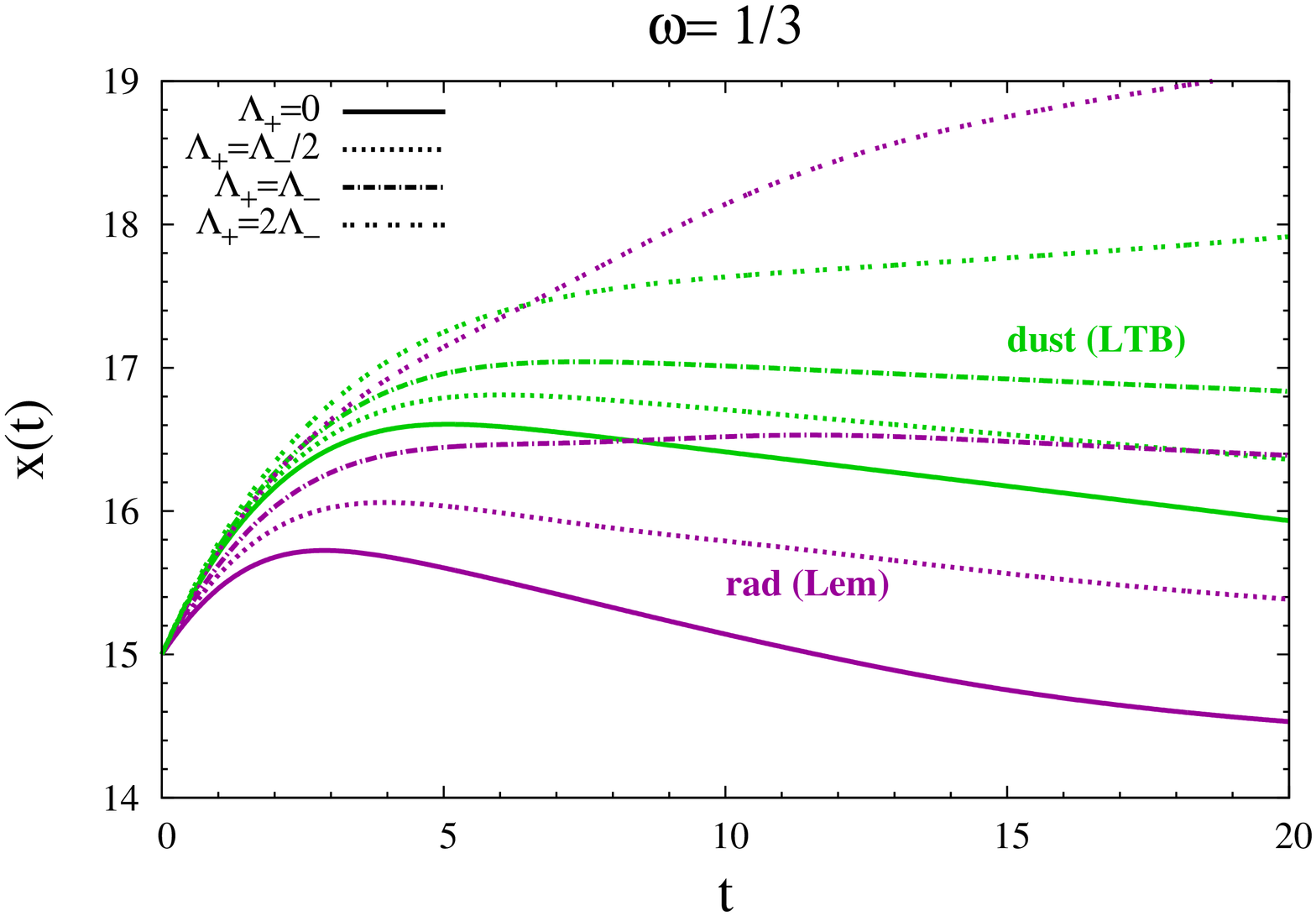}
\caption{Comparison between the evolution in inhomogeneous radiation backgrounds and inhomogeneous dust background. The geometry of the external region is described, respectively by Lema\^{\i}tre's and the LTB solution.  We choose for our examples $r_0=20$, $s_0=5$ and $\delta_{\epsilon}=0.5$.} 
\label{fig:LTBvsLem}
\end{figure}

  Figures~\ref{fig:LemVsrad} and \ref{fig:LTBvsLem} show the growth of the external radial coordinate of the bubble considering $w=0$ and $w=1/3$. In figure~\ref{fig:LemVsrad} each curve is compared with the evolution in the corresponding homogeneous radiation background, while in figure~\ref{fig:LTBvsLem} the evolution in LTB and Lema\^{\i}tre's backgrounds are shown together. 
  The curves in Fig.~\ref{fig:LemVsrad} show that the bubble grows slower in the inhomogeneous radiation case (when compared to the case with homogeneous radiation), while those in Fig.~\ref{fig:LTBvsLem} show that the evolution of the bubble is slower in the case of inhomogeneous radiation (compared to that of inhomogeneous dust). In both figures, the only exception is the case 
  $\Lambda_+ = 2\Lambda_-$, due to the more rapid dilution of the background density for radiation. 

\begin{figure}[tbp]
\centering 
  \includegraphics[angle=-90,width=7.5cm]{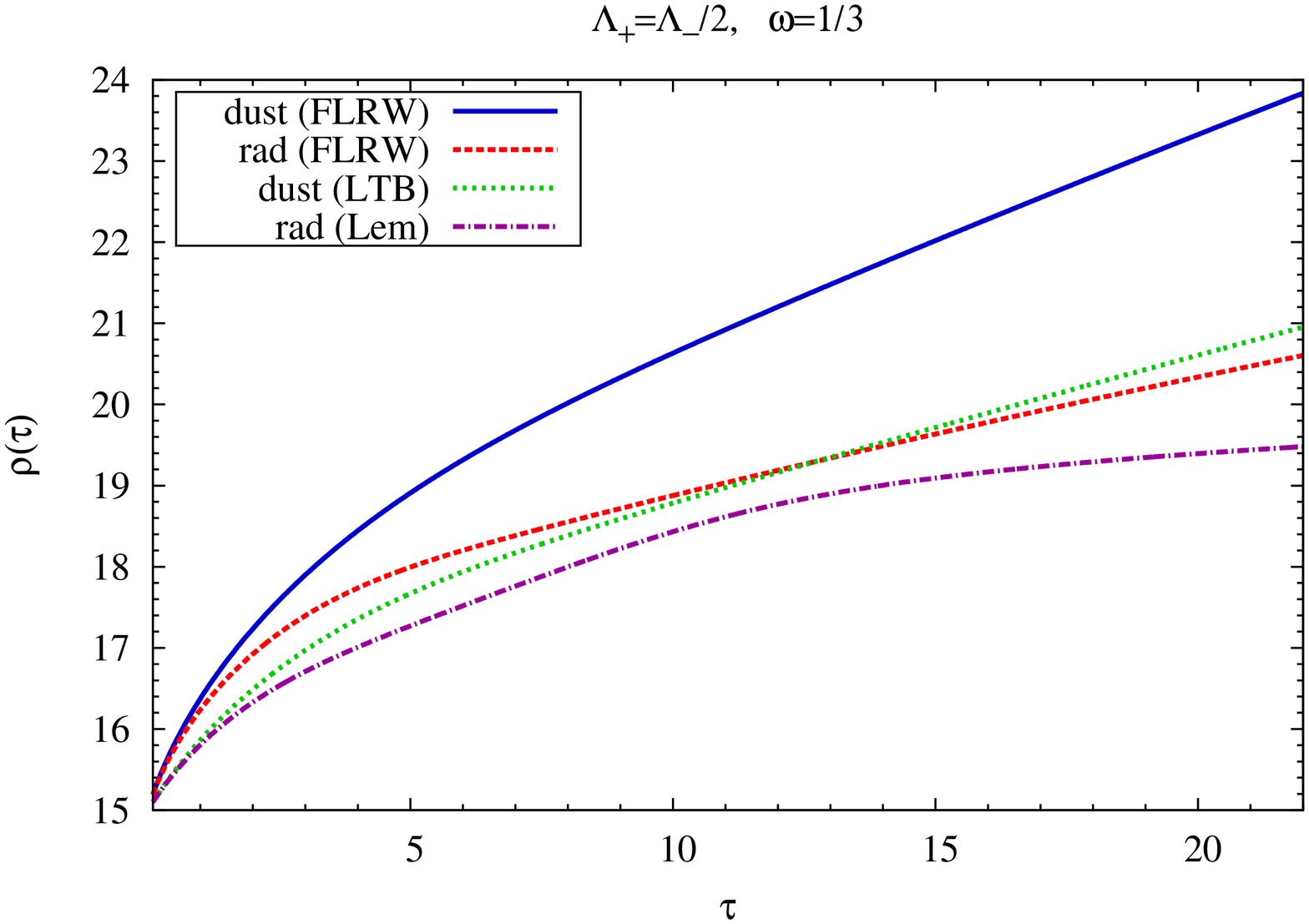}
  \hfill
  \includegraphics[angle=-90,width=7.5cm]{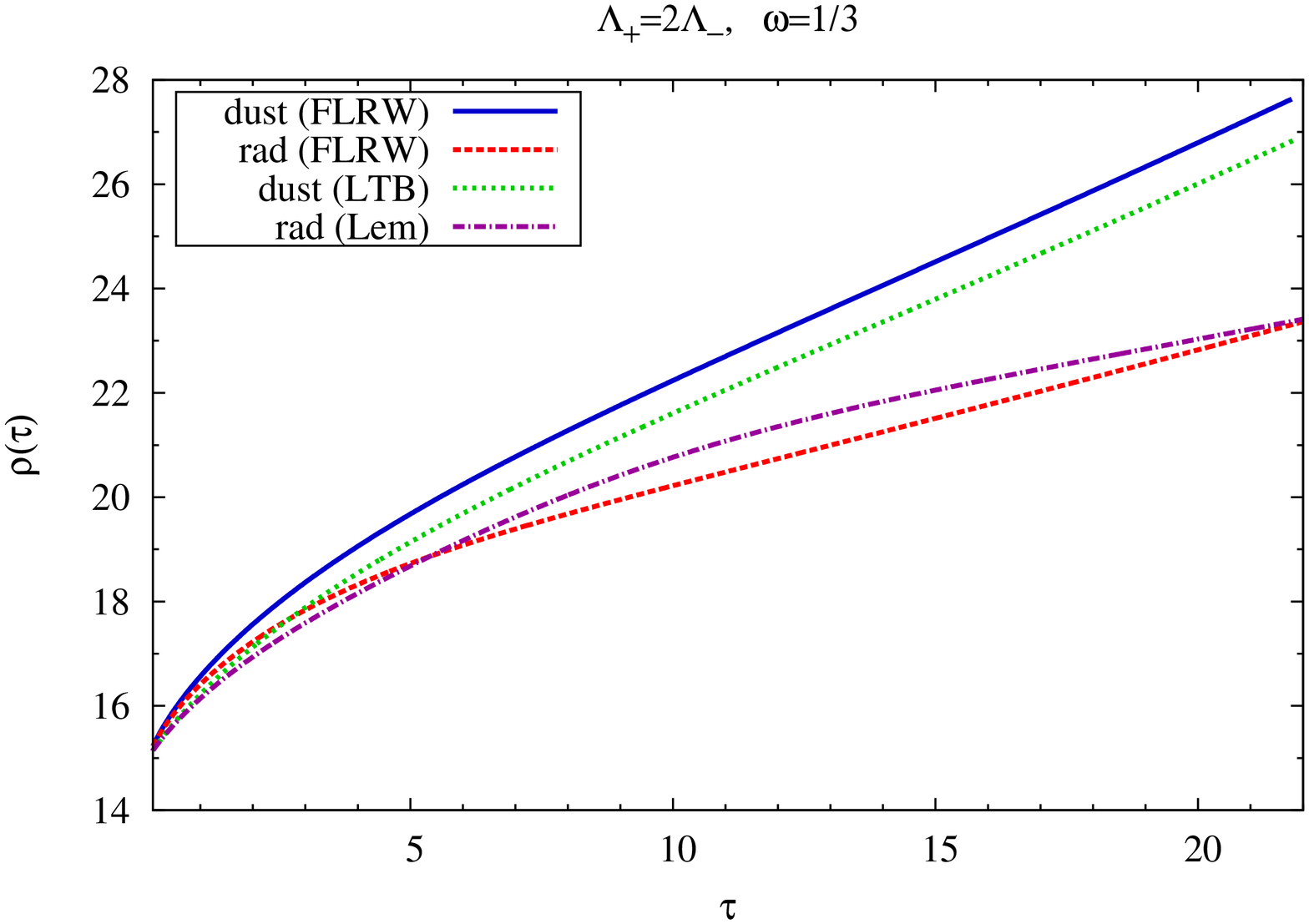}
  \caption{Evolution of the proper radius of the bubble expressed in terms of the proper time for different outer space-times. In those cases in which the bubble is in a radiation ambient, the growth of $\rho$ is slower than in the corresponding dust case.} 
\label{fig:ProperRadius}
\end{figure}

  The dependence with the parameter $\Lambda^+$ can be also analysed by considering the evolution of the proper radius of the bubble, as displayed in figure~\ref{fig:ProperRadius}. We observe that the evolution of the bubble is noticeably affected by the background in the following aspects: (i) in those cases in which the bubble is in a radiation ambient, the growth of the proper radius is slower than in the corresponding dust case, and (ii) the evolution depends on the radial distribution of the outer radiation, as well as on the value of the outer cosmological constant.
  
%
%
%
\section{General Discussion}
\label{sec:discussion}
We presented a study of the evolution of vacuum bubbles in backgrounds with inhomogeneous dust or radiation matter content, and compared it with the corresponding homogeneous cases. This analysis is important in the context of inflationary models, as a first step in the description of the growth of vacuum regions in the presence of inhomogeneities  generated during a pre-inflationary era, and the influence of these in such a growth.

 We have developed a numerical code to compute the evolution of vacuum bubbles using the thin-shell formalism. The problem involves the integration of a system of partial differential equations to determine the evolution of the radial coordinate and the energy density of the bubble, together with the evolution of the geometry of the background. This geometry is described by the FLRW metric in homogeneous cases, and the LTB and Lema\^{\i}tre's metrics for inhomogeneous dust and radiation cases, respectively. Our code reproduces those results for cases with dust backgrounds previously obtained by other authors \citep{Fischler2008,Simon2009}, and also generalises the problem to those evolutions in radiation backgrounds (with both homogeneous and inhomogeneous distributions). 

 We have computed the evolution for different values of the parameters $\Lambda^+$ and $w$, which characterise the external geometry and the matter content of the bubble, respectively. The comparison between cases with homogeneous backgrounds of dust or radiation, described by the FLRW metric, shows that 
  the radiation content in the external region slows down the evolution of the bubble, as long as it is not governed by $\Lambda_+$.
The analysis of the inhomogeneous cases shows that the evolution is initially delayed, when compared with the corresponding homogeneous cases, if the bubble nucleates in a sub-density region. 
Regarding to the matter content of the bubble, although evolutions for the values $w=0$ and $w=1/3$ are qualitatively similar, a bubble with a radiation content expands faster than a bubble with a dust content. 
Notice also that the evolution is monotonic only in the case $\Lambda^+=2\Lambda^-$ in the presence of radiation.

  It is important to emphasise that $x(t)$ represents the radial external coordinate of the bubble and indicates the growth of the bubble with respect to the non-comoving expanding background, so the slowing down of $x(t)$ must not be interpreted as a collapse scenario. The evolution of the proper radius of the bubble (shown in  figure~\ref{fig:ProperRadius}) is affected by the  background features in the following aspects: (i) the growth of the proper radius is slower in a radiation ambient than in the corresponding dust case, and (ii) the evolution depends on the both the radial distribution of the outer radiation and the value of the outer cosmological constant.
  
There are several possibilities for extensions of our work. Among them we shall mention three. 
First, the setting used here can be applied to the eternal inflation scenario, with the appropriate initial conditions, namely those that are not contaminated by unrealistic decaying modes which diverge as $t\rightarrow 0$.\footnote{We thank an anonymous referee for this remark.} 
This could be done along the lines of refs.~\cite{Shibata1999,Polnarev2007}.
Second, it would be interesting to develop a more detailed study of the dependence of the evolution of the bubble with the initial profiles, to assess the issue of genericity of inflation in inhomogeneous backgrounds. Third, the evolution in different backgrounds may leave signatures in the inflating region. Since the bubble plays the role of a moving boundary of this region and, as we have shown, the presence of inhomogeneities outside the bubble modifies its motion, quantum fields  inside the bubble will be indirectly influenced by the external inhomogeneities. 
We hope to return to these issues in future publications.

\appendix
\section{Lema\^{\i}tre's geometry}
\label{appLEM}

\begin{figure}[tbp]
\centering 
  \includegraphics[width=7.6cm]{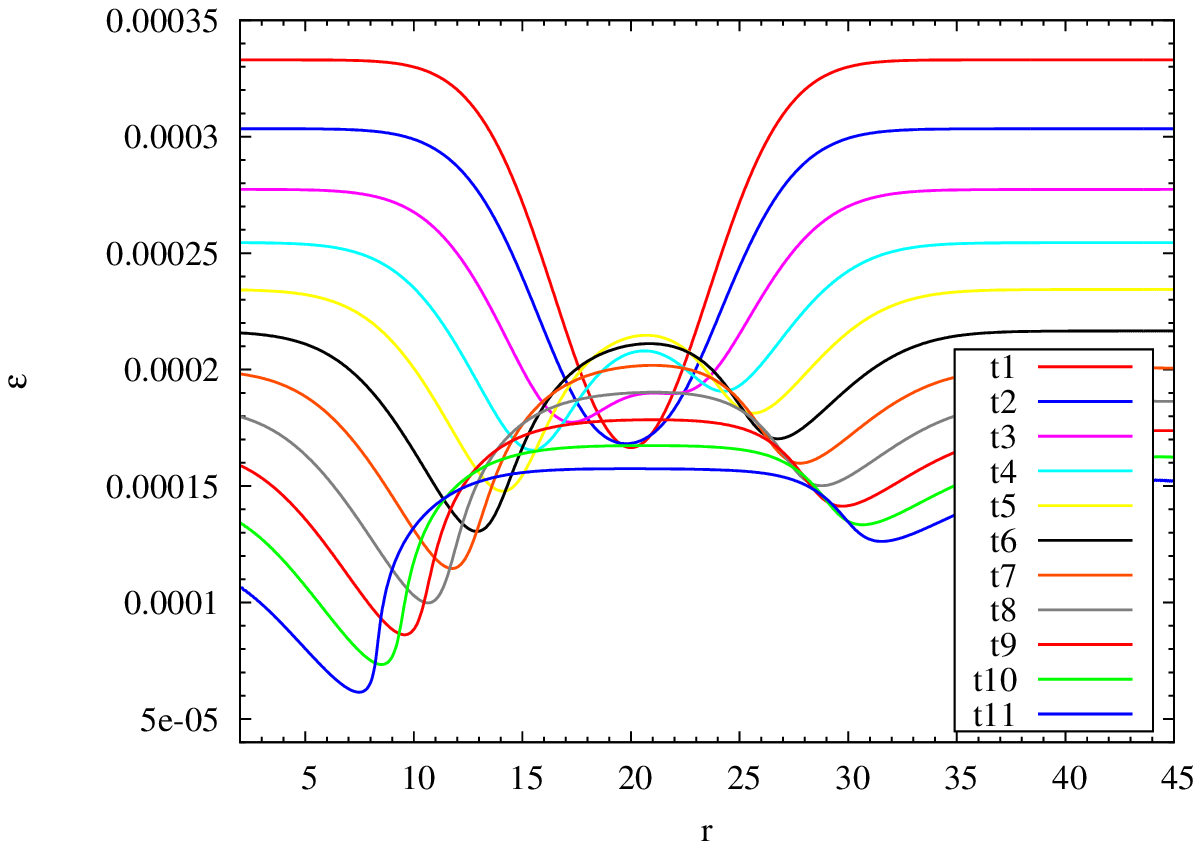}
  \hfill
  \includegraphics[width=7.6cm]{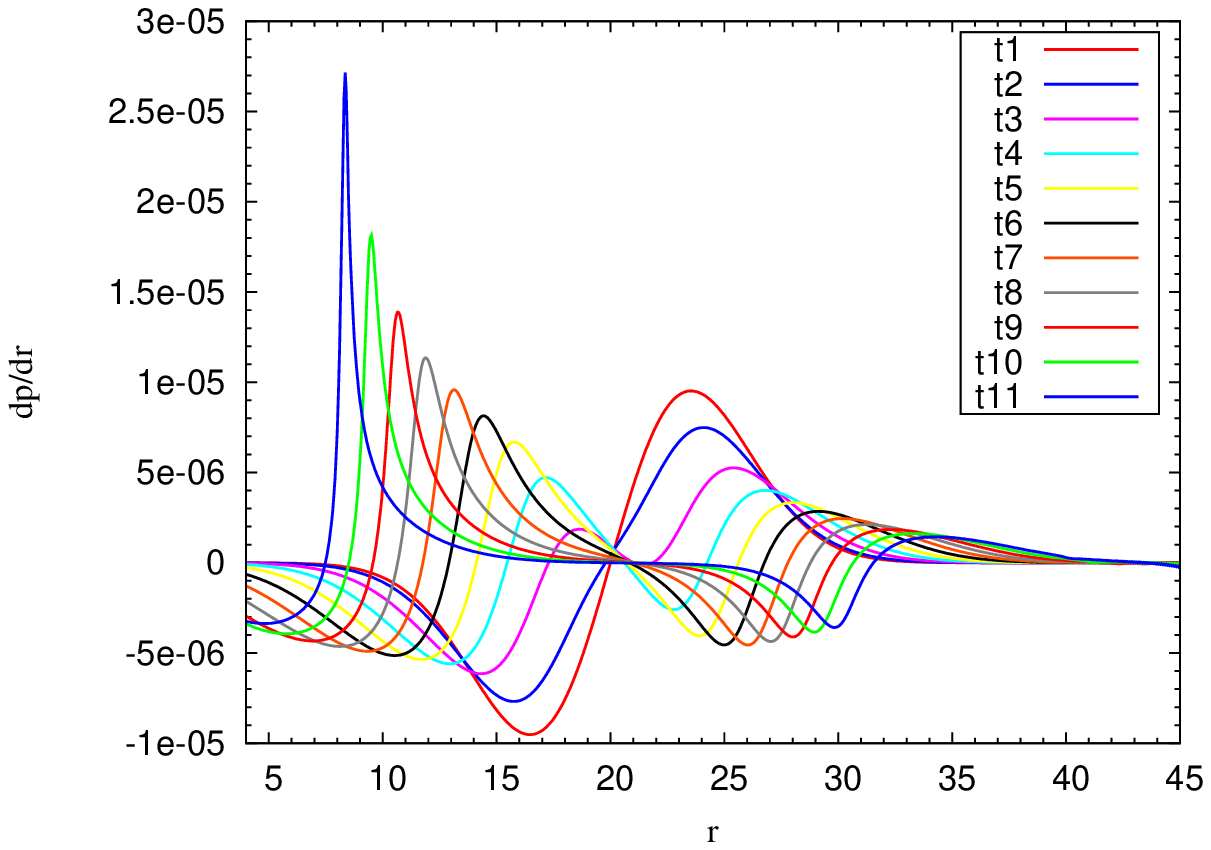} \\ 
  \vspace{.5cm}
  \includegraphics[width=7.6cm]{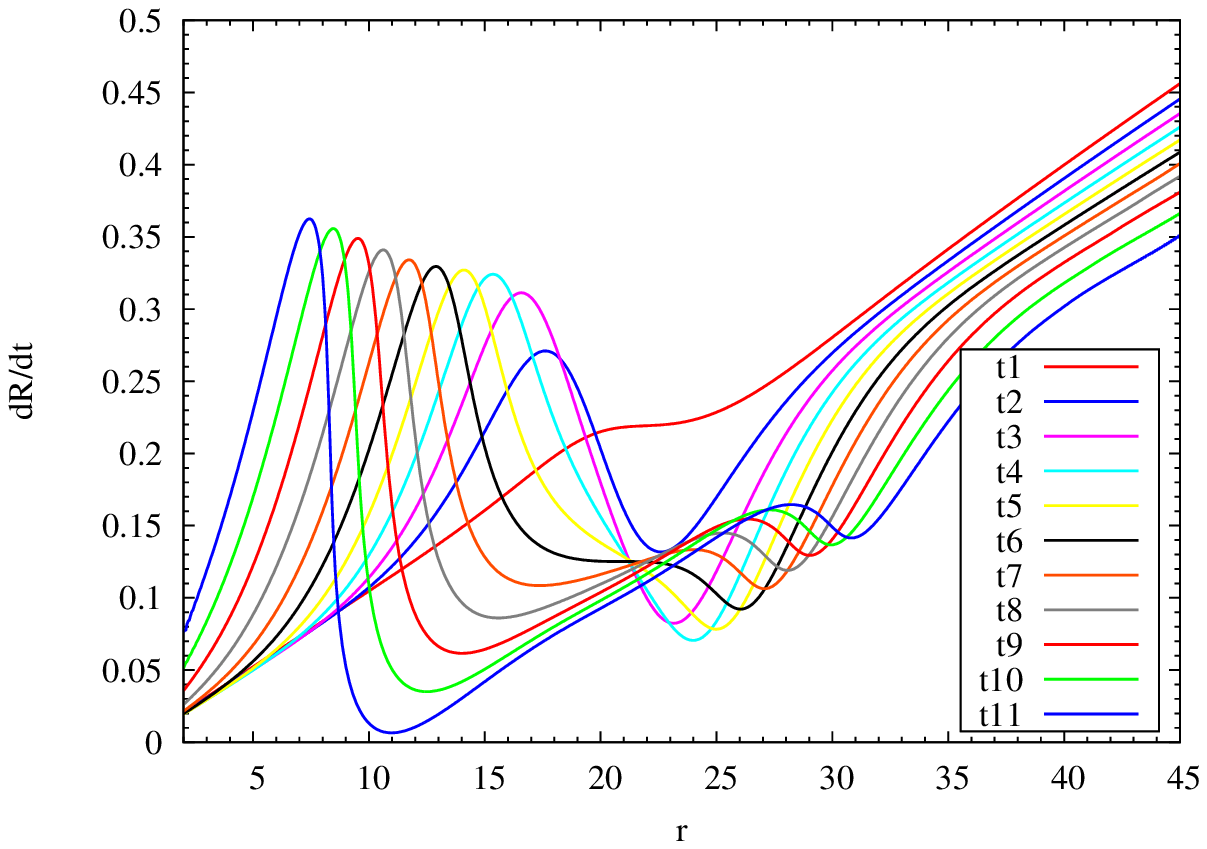}
\caption{Evolution of the functions which characterise Lema\^{\i}tre's geometry, namely the density, the pressure gradient and the expansion rate.  The initial conditions are those described in section \ref{sec:burbuja_num} with $\Lambda^+=0$. The labels $t_i$ indicates successive time steps of the evolution. A pronounced pressure gradient causes the propagation of the inhomogeneous region away from the corresponding initial radial coordinate. If $\dot{R}$ eventually changes its sign, the geometry will represent  a collapsing scenario.}
\label{fig:Lemaitre}
\end{figure}

The features present in the evolution of Lema\^{\i}tre's solution can be qualitatively understood following the discussion in ref.~\cite{Bolejko2008}. Let's consider the evolution Eq.~(\ref{dRdt}), which can be rewritten as 
\begin{equation}
   {\rm e}^{-A}\dot{R}^2=\frac{2M}{R}+\frac{1}{3}\Lambda^+ R^2 -1 +(1+2E){\rm exp}\left(-2\int {\rm d}t\frac{p'}{(\epsilon+p)} \frac{\dot{R}}{R'}\right)\,.
\end{equation} 
The l.h.s. is associated to the expansion rate of the external space-time.
In regions in which the initial profiles are such that the pressure gradient is large, the exponential will decrease and the expansion rate of shells with $r={\rm constant}$ will be reduced. Relative to these, shells with larger values of $r$ will expand faster, leading to a drop in the gradient of $p$, and eventually to a change of sign in $p'$. Negative values of $p'$ cause the increment of the expansion rate, hence leading to acoustic oscillations, which were previously analysed in ref.~\cite{Bolejko2008}, and are noticeable in figure~\ref{fig:Lemaitre}. If the oscillations grow enough to change the sign of $\dot{R}$, then a collapse of the geometry could take place at different radial coordinates. 

Unlike the LTB solution, the inhomogeneous regions are not confined to a fixed initial radial coordinate. This behaviour of the evolution of the geometry is a direct consequence of the not-zero pressure gradient which characterises Lema\^{\i}tre's solution (note that in the particular case with $p'=0$, the metric functions in Eqs.~(\ref{ds_Lem}) reduce to the form $A(t,r)=0$ and ${\rm e}^{B(t,r)}=\frac{R'(t,r)}{(1+2E(r))}$, that is, the LTB limit is recovered).

\acknowledgments

 FATP acknowledges support from CONICET and CLAF/ICTP. SEPB would like to acknowledge support from FAPERJ and UERJ. 


\bibliographystyle{JHEP} 
\bibliography{bibliography} 

\end{document}